\documentclass[prd,nofootinbib,twocolumn]{revtex4}
\usepackage{amsmath}
\usepackage{amssymb}
\usepackage{graphicx}
\usepackage[usenames, dvipsnames]{color}
\newcommand{\red}{\textcolor{red}}

\newcommand{\Epeak}{\dot{E}_{\rm peak}}
\newcommand{\Mrem}{M_{\rm rem}}
\newcommand{\Jrem}{J_{\rm rem}}
\newcommand{\chirem}{\chi_{\rm rem}}
\newcommand{\Jdot}{\Dot{J}}
\newcommand{\Edot}{\Dot{E}}
\newcommand{\OmegaH}{\Omega_{\rm H}}
\newcommand{\Lin}{L_{\rm in}}
\newcommand{\Jrad}{J_{\rm rad}}
\newcommand{\Jspin}{J_{\rm spin}}
\usepackage{graphicx}

\newcommand{\Lk}{L_{\pm}^{\rm K}}

\usepackage{amsmath}
\usepackage{graphicx}
\usepackage[colorlinks=true, allcolors=blue]{hyperref}

\begin{document}

\title{Binary Black Hole Coalescence Phenomenology from Numerical Relativity} 
\author{Richard H.~Price}
\affiliation{Department of Physics, MIT, 77 Massachusetts Ave., Cambridge, MA 02139}
\author{Ritesh Bachhar}
\affiliation{Department of Physics, East Hall, University of Rhode Island, Kingston, RI 02881}
\author{Gaurav Khanna}
\affiliation{Department of Physics and Center for Computational Research, University of Rhode Island, Kingston, RI 02881}
\affiliation{Department of Physics and Center for Scientific Computing \& Data Research, University of Massachusetts, Dartmouth, MA 02747}


\begin{abstract}
\noindent 
The major source of ground-based gravitational wave  detectors, the inspiral and merger of comparable mass binary black holes (BBH), consists of a slow quasicircular inspiral, a merger to form a single remnant hole, and the quasinormal ringing of that remnant. The first and last of these epochs are amenable to well developed and familiar approximations: Newtonian or post-Newtonian  for the first, and BH perturbation methods for the last. Ironically, the middle epoch, the merger, generates most of the GW emission, yet has been accessible so far only to numerical relativity. Here we add the close-limit approximation for the phenomenology of the merger. With the completed set of methods (Newtonian/post-Newtonian; close limit; BH perturbation theory) we show that not only can we understand the results of  BBH coalescence, but -- with reasonable accuracy -- we can {\it predict} the resulting radiation and remnant.


\end{abstract}

\maketitle
\section{Introduction and Overview}

Computational modeling of binary black hole coalescence has evolved to a well-established field driven by the data analysis requirements of gravitational wave detectors such as LIGO. Prior to the middle of the first decade of this millennium our understanding of the process, and indeed most of black hole astrophysics, depended on approximation methods, in particular the linearization of Einstein's equations about a flat background or about the background of the final merged black hole. The dramatic progress, since then, in numerical simulations (hereafter NumRel)\cite{BaumShap} of the process involving the fully nonlinear Einstein equations 
can be inferred from the fact that there is now a catalog~\cite{SXScatalog,SXSpaper} of results of these simulations representing (as of this writing) 2,265 models. 

These numerical results are as important as they are impressive, but for some they do not give a fully satisfactory heuristic or phenomenological picture of the physical processes involved. Such pictures can be useful for suggesting new directions as well as for outreach. In this paper we attempt to address this with a two-pronged approach: First, we will deconstruct a NumRel model and relate its details to physics. Second, we will investigate the extent to which a physical picture can be combined with a mixture of approximations to create a NumRel-independent quantitative model of the coalescence that may be sufficient for some purposes.

We summarize here  certain abbreviations that we will use
throughout the paper. BH=black hole; BBH=binary black hole; NumRel=numerical relativity; GW=gravitational wave; GR=general relativity, Einstein's theory of gravity; $L$ will represent orbital angular momentum while $J$ stands for angular momentum more generally; ``rem," as a subscript or superscript, will indicate the remnant merged hole, so that, e.g., $\Jrem$ is the angular momentum of the remnant; the Kerr parameter $a$ will represent the spin of a BH divided by its mass; $\chi$ will be the spin parameter, $a$ divided by BH mass. The masses of the individual BHs will be $m_1,m_2$ with $m_1\geq m_2$ and $M=m_1+m_2$ will represent the total mass-energy of the system; we use the common notation  $q\equiv m_1/m_2$. As is typical in GR work we take $c=G=1$. Dimensional quantities will be normalized by taking the initial value of the system mass $M$ to be unity. 

What we will be plotting, and calling $\Omega(t)$ will be inferred from NumRel simulations of the GW waveforms. What we shall be plotting is {\it half}~\footnote{The factor of 1/2 can be understood as a consequence of the quadrupolar nature of the GWs.} the GW frequency. In general this will be treated as the orbital angular velocity of the pre-merger BHs.

To describe the stages of the binary black hole process we shall adapt the vocabulary introduced in Ref.~\onlinecite{vocab}.  The early stage of the binary's evolution, the stage at which there are two distinct objects losing energy to outgoing GWs, will be the ``inspiral." The inspiral stage  ends with the ``merger" stage, when the BHs lose their individual identity, merging into a single object that is initially highly dynamical and involving relativistically strong gravitational fields. The violent dynamics of the merger subsides and transitions to ``ringdown," damped oscillations at specific frequencies with specific damping. The damped ringing leaves a single quiescent BH, the ``remnant."  The entire binary life story is the ``coalescence."

BBH coalescence models are characterized by the ratio of BH masses, $q$, the strength of each hole's spin $\chi$, and the orientation of that spin. Too wide a set of models might obscure  a relatively simple connection we hope to make between NumRel results and the physical phenomenology of the binary process. For that reason we will limit ourselves to two models, both with $q=1$ and which involve holes with the same $\chi$. The first model will have $\chi=-0.8$ (high rate of spin, antialigined with orbital angular momentum) for each of the holes; in the second  model, the holes will have no spin ($\chi=0$). For compactness of notation we will refer to these below as our $\chi=-0.8$ and $\chi=0$ models. The first choice is made to illustrate that our phenomenology is not restricted to the spinless case; the second choice is needed to illustrate the success of the close limit approximation (CLAP) for which work remains to be done for BHs with spin.

The rest of this paper is organized as follows. Section~\ref{sec:PhenFromNumRel} presents results from NumRel simulations: waveforms, $\Omega(t)$, the rate of outgoing GW energy $\Edot$, and of angular momentum $\Jdot$. These will be viewed from the three-stage (inspiral, merger, ringdown) picture of BBH coalesence. It will be shown that simple models of the pre-merger and post-merger stages give excellent agreement with the NumRel simulations. The relatively short merger stage is the focus of Sec.~\ref{sec:peakEdot}. To deal with that short stage, in that section  we introduce an innovative application of the close limit approximation, the ``CLAP"~\cite{EdotVsChi,CLAP1,CLAPcomparison,CLAP2,CLAP3,AndradePrice,SopuertaYunesLaguna}. In Sec.~\ref{sec:NumRelInd} we turn from the physical interpretation of the NumRel results to the question of whether, or to what extent, such results can be predicted from approximation methods.

\section{Pre- and post-merger phenomenology inferred from NumRel GWs }\label{sec:PhenFromNumRel}

\begin{figure}[h]
    \centering
    \includegraphics[width=.55\textwidth]{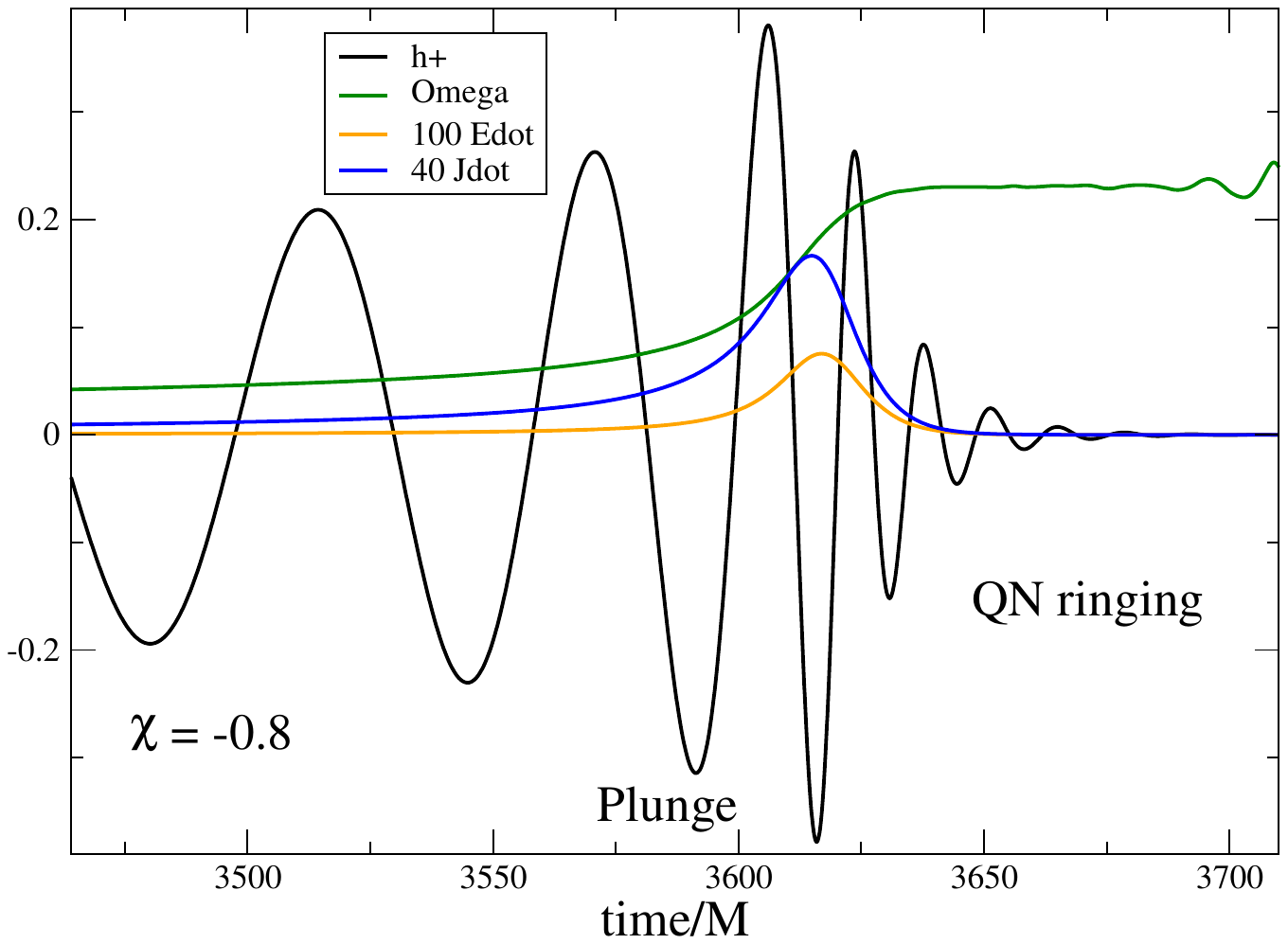}
    \caption{Primary quantities as functions of time for a BBH with $q=1$ and initial BH spins $\chi=-0.8$ anti-aligned with orbital angular momentum.}
    \label{fig:introplot}
\end{figure}

\begin{figure}[h]
    \centering
    \includegraphics[width=.55\textwidth]{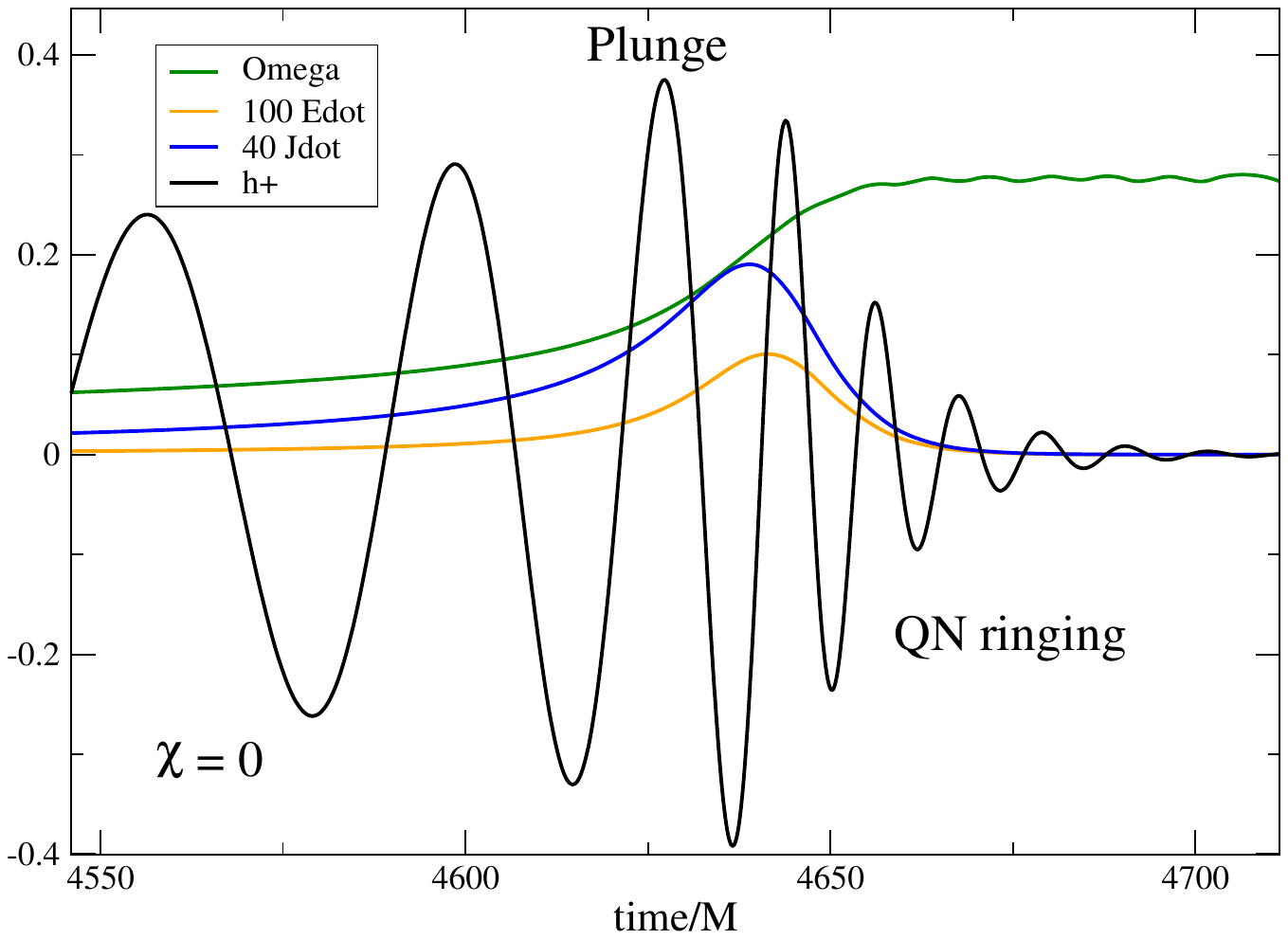}
    \caption{Primary quantities as functions of time for a BBH with $q=1$ and initial BH spins $\chi=0$ anti-aligned with orbital angular momentum.}
    \label{fig:introplot2}
\end{figure}

Figure~\ref{fig:introplot} gives the primary results for the model with $q=1$ and $\chi=-0.8$ for each hole.  This model is SXS:BBH:0154 in the SXS catalog.~\cite{SXScatalog}
 For clarity of the figure, $\dot{E}$, the rate of emission of gravitational wave energy, is multiplied by 100, and the rate of angular momentum emission $\dot{J}$, in units of the total mass $M$, is multiplied by 40. Figure~\ref{fig:introplot2}
shows the same quantities for $\chi=0$. Since the primary features in this figure are the same as for those in Fig.~\ref{fig:introplot}, for definiteness we will limit our discussion only to the $\chi=-0.8$ case of Fig.~\ref{fig:introplot}.

The waveform in Fig.~\ref{fig:introplot} shows the progression of the binary through the stages of the coalescence. Before $t/M\approx3600$ the inspiralling binary evolves in quasi-circular orbits with gradually increasing GW amplitude and frequency. In the particle perturbation approach the particle, gradually losing energy to GW emission, would reach an innermost stable circular orbit, the ``ISCO," followed by a plunge of the particle. For comparable mass black holes there can be no strict equivalent of an ISCO, since -- unlike the case of a perturbing particle -- there is no conserved energy, but the concept is parallel: There comes a point at which the system no longer supports quasi-circular inspiralling  and the BHs, in a sense, plunge into each other; the system merges.

In Fig.~\ref{fig:introplot} this can be seen to happen at a time around 3600\,$M$. The damped sinusoids of ringdown of the remnant clearly start by $t/M=3640$, and perhaps as early as 3620. The time roughly from 3600\,$M$ to 3620\,$M$ is the merger,  the period during with most of the energy and angular momentum of the process are radiated, and a main focus of this paper.

\begin{figure}
    \centering
    \includegraphics[width=.45\textwidth]{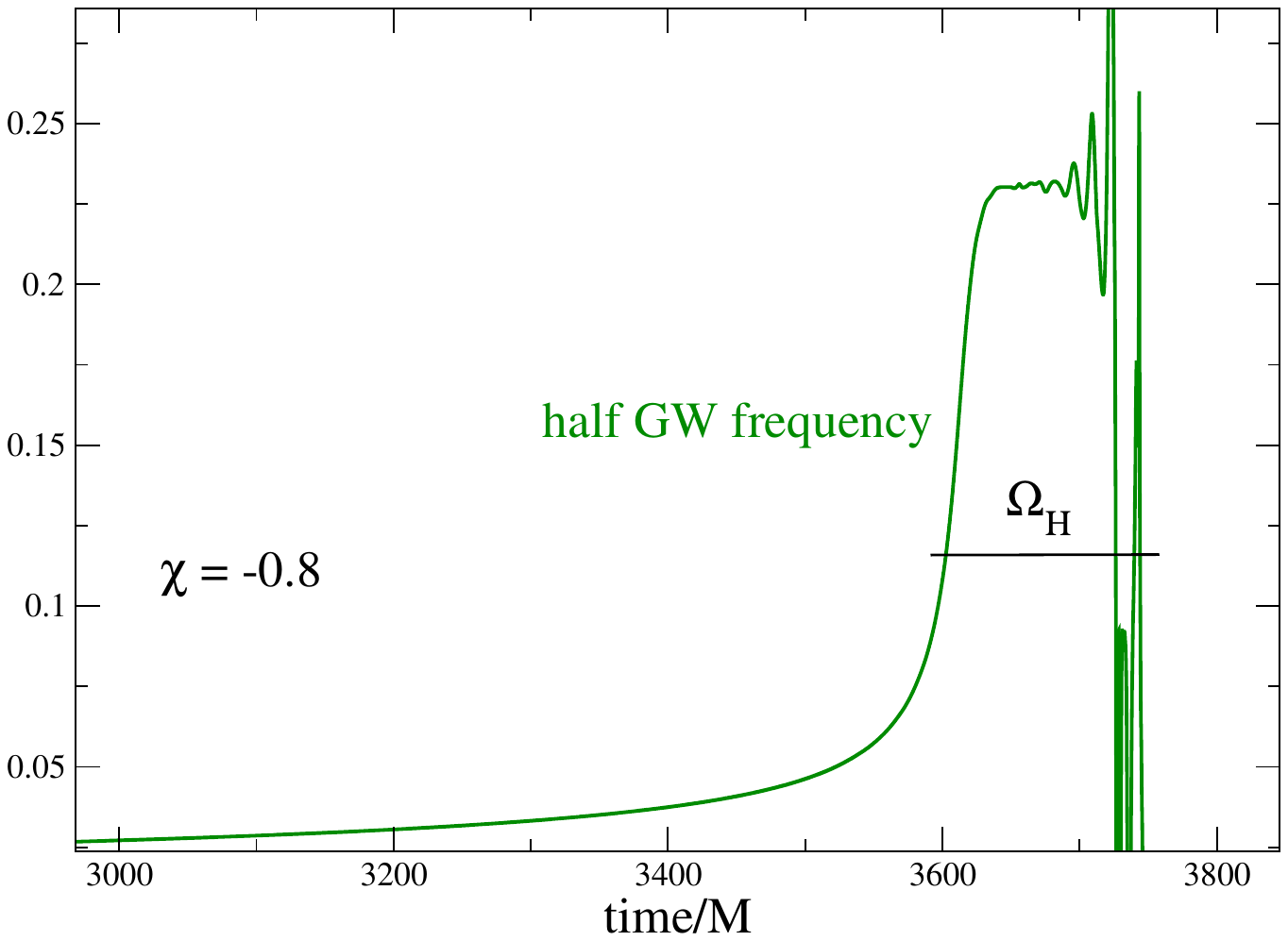}
    \caption{Behavior of $\Omega$ in the three epochs of inspiral.}
    \label{fig:justomega}
\end{figure}

In the history of the process, it is useful to focus on $\Omega(t)$, as is done in Fig.~\ref{fig:justomega}. 
It is particularly interesting to look at the sudden rise in $\Omega$ between $t/M$=3600 and 3635.  
 Figure~\ref{fig:introplot} shows that the time duration of this rise is only a single GW oscillations, emphasizing the care needed in interpreting $\Omega$ (half the GW frequency) as the orbital frequency.

Particularly misleading in $\Omega(t)$ is  the constant-$\Omega$ plateau from about $t/M=3635$ to 3665, a plateau that follows the merger. The curve might suggest constant orbital frequency as in the ISCO orbiting of a particle. It is, however, the QN ringing at a complex frequency $M^{\rm rem}\omega= \omega_R+\omega_I=0.4427-i\, 0.08675$ for the remnant Kerr hole with spin parameter $a=0.41254M$ of this model. For this damped ringing, the GW oscillations, at frequency $~0.44/M$, agrees with the plot in Fig.~\ref{fig:introplot} showing $\Omega$, {\it half} the GW frequency around 0.22/M.

Orbital frequency  has no straightforward meaning after the merger results in a single rotating hole. One might wonder whether the extension of orbital frequency might be the rotation rate of the remnant hole. For a Kerr hole of mass $M$ and  spin parameter $a$ this is often taken to be the horizon rotation rate~\cite{Poisson70,Poisson80}
\begin{equation}
    \Omega_{\rm H} = \frac{a}{2M(M+\sqrt{M^2-a^2\;} )}\,.
\end{equation}
We apply this to the model by taking $M\OmegaH=\chi_{\rm rem}/(1+\sqrt{1-\chirem^2}\ )$, where $\chirem$ is the remnant angular momentum for the model divided by the square of the remnant mass.
This value, for our model, is indicated in Fig.~\ref{fig:justomega}.

\begin{figure}
    \centering
    \includegraphics[width=.55\textwidth]{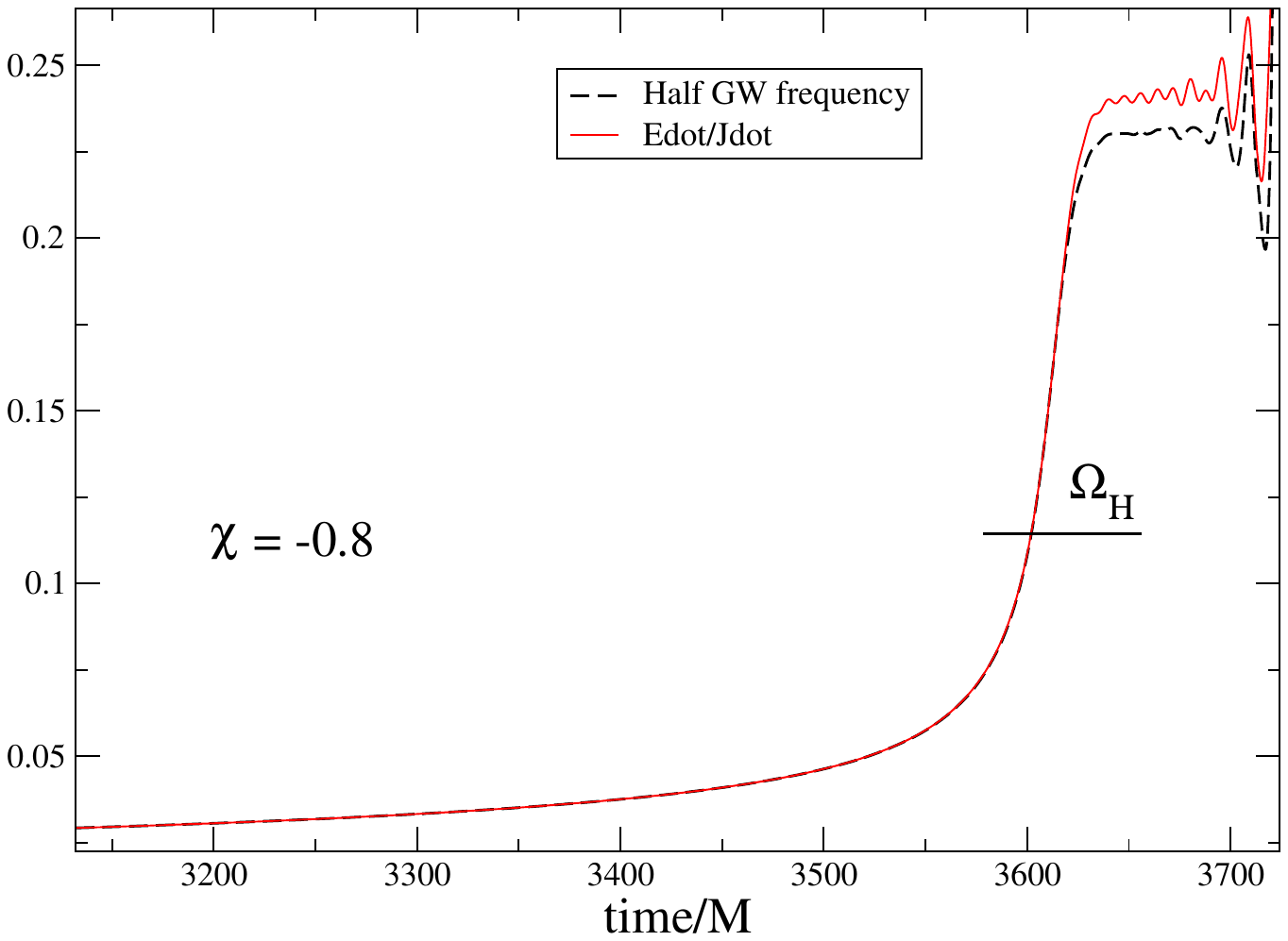}
    \caption{$\dot{E}/\dot{J}$ compared with $\Omega$, the GW frequency.}
    \label{fig:EbyJ}
\end{figure}

It is interesting now to correlate the stages of the coalescence with the pattern of the GWs.
The curves shown in Fig.~\ref{fig:introplot} are for a particular BBH inspiral, but all models we have looked at, show the same qualitative features. For the most part we will be interested in the curves for $\dot{E}$ and $\dot{J}$. These emissions are closely related. Though $\dot{J}$ in our results is computed from the GWs in the radiation zone, for the inspiral stage it is well approximated by $\dot{E}=\Omega\dot{J}$, where it should be remembered $\Omega$ is half the frequency of the GW emission. The relationship is displayed in Fig.~\ref{fig:EbyJ}, which shows that the   $\dot{E}=\Omega\dot{J}$   relationship 
breaks down, as expected, in the merger. We note also, that there is no apparent relevance of $\OmegaH$.

\subsection{Approximations near the plunge}

\subsubsection{Pre-merger fit}

The rate of GW energy emission according to the quadrupole formula for Newtonian point masses is the well-known 1963 Peters-Matthews formula~\cite{PetersMatthews,MTW,MTWquadform}
\begin{equation}\label{eq:PetersMatthews}
    \dot{E}=\frac{32}{5}\frac{m_1^2m_2^2M}{R^5}\,,
\end{equation}
where $M=m_1+m_2$ is the system mass.
Newtonian dynamics can then be used for the relationship of orbital angular velocity $\Omega$ to point mass separation $R$ to convert Eq.~\eqref{eq:PetersMatthews} to

\begin{equation}\label{eq:PetersMatthews2}
   \dot{E}=\frac{32}{5} \frac{M^{10/3}\Omega^{10/3}}{(1+q)^2(1+1/q)^2} .
\end{equation}

\begin{figure}
    \includegraphics[width=.45\textwidth]{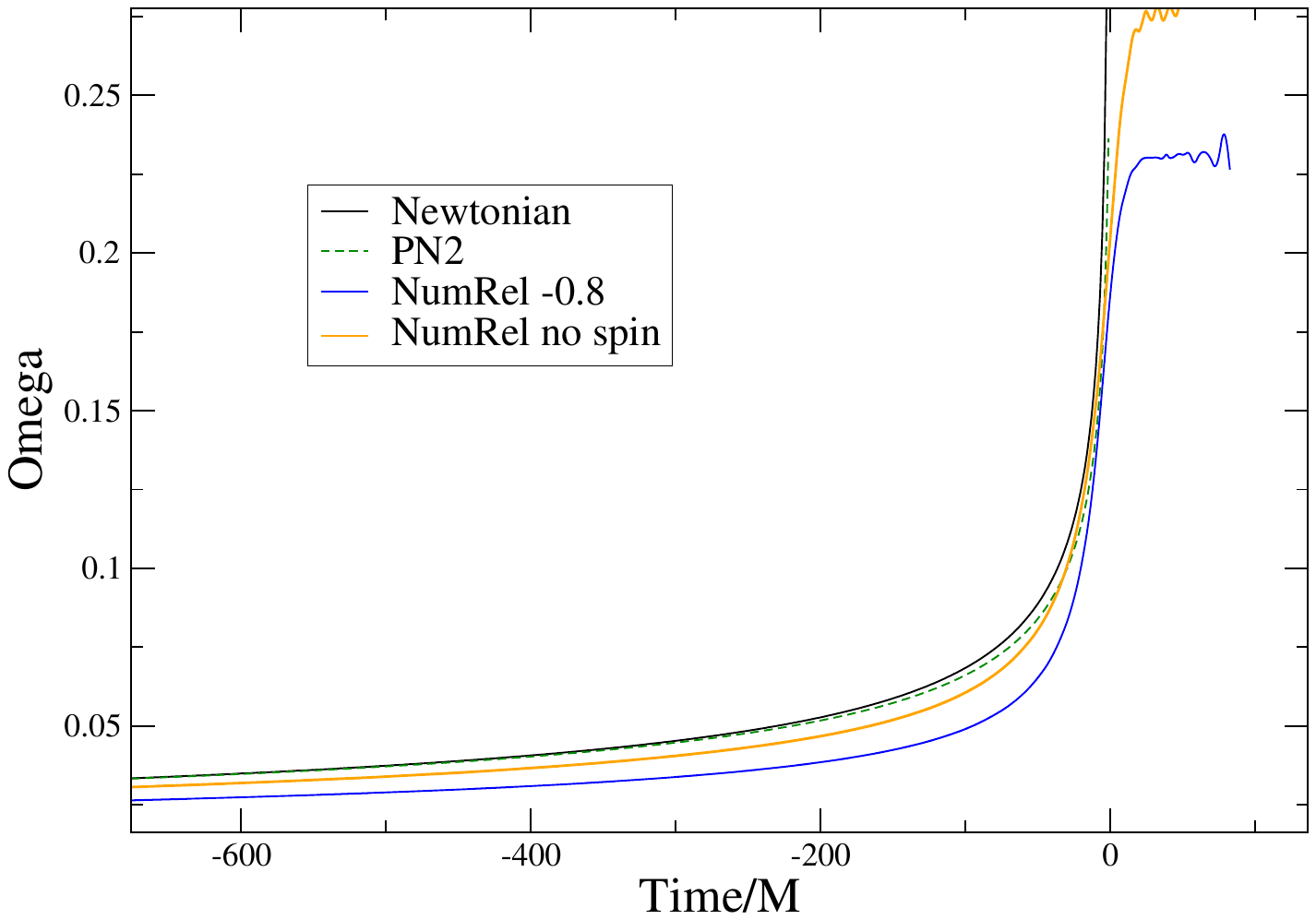}
    \caption{Comparison of $\Omega(t)$ for $q=1$ symmetric models. 
   The continuous (black) Newtonian curve follows from  Eq.~\eqref{eq:PetersMatthews2} and from the Newtonian expression for the energy of the equal mass binary $-(1/8)M^{5/3}\Omega^{2/3}$.
    The dashed (green) curve is directly from Eq.~(8) of the second-order post-Newtonian (PN)analysis of Ref.~\onlinecite{BlanchetEtAl}. NumRel results are shown for  the  $\chi=0$ and $\chi=-0.8$ models. All curves have been time-shifted to put the time of coalescence at $t=0$.} 
    \label{fig:PN2rvsdNumRel}
\end{figure}

For a symmetric binary this simplifies to $\Edot=(2/5)M^{10/3}\Omega^{10/3}$. By combining this $\Edot$ with the Newtonian orbital energy $-(1/8)M^{5/3}\Omega^{2/3}$, we can infer the Newtonian (plus quadrupole formula) result $\Omega(t)=(1/8)((-t)/20)^{-3/8}$, where we are choosing $t=0$ as the time of coalescence. This result is displayed in Fig.~\ref{fig:PN2rvsdNumRel}, along with the $\Omega(t)$ from the second-order post-Newtonian approximation\cite{BlanchetEtAl},   and the $\Omega(t)$ from NumRel.
The reasonably good agreement of these curves before the merger is not surprising. The speed of one of the BHs ``particles" in the Newtonian approximation is $(GM)^{1/3}\Omega^{1/3}/2$, or simply $\Omega^{1/3}/2$ in our $G=M=1$ units.  At $\Omega\approx0.2$, therefore, the speed (in the Newtonian approximation) approaches  only 10\% of $c$.

\begin{figure}
    \centering
    \includegraphics[width=.45\textwidth]{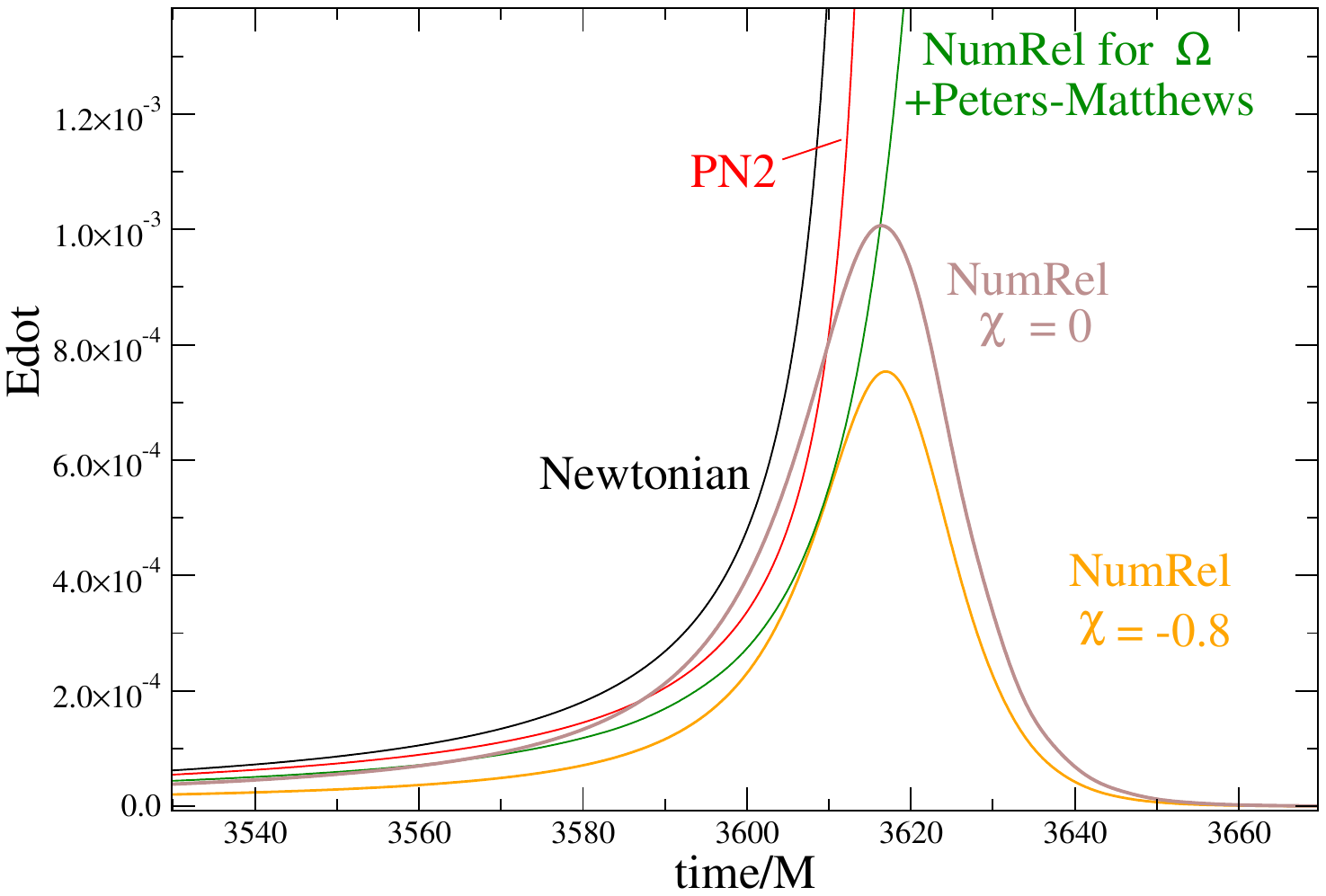}
    \caption{The $\dot{E}$ NumRel result for our models compared with the approximation using Eq.~\eqref{eq:PetersMatthews2} applied both to the Newtonian and PN approximations to $\Omega(t)$. The Newtonian and PN curves have been shifted in time for the best fit to the NumRel curve at early times.
    }
    \label{fig:preplunge}
\end{figure}

We now  apply  Eq.~\eqref{eq:PetersMatthews2} to the Newtonian and PN approximations for $\Omega(t)$ to get  estimates of the pre-merger $\Edot$, and show the results in Fig.~\ref{fig:preplunge}.
That figure also shows the $\Edot$ from Eq.~\eqref{eq:PetersMatthews2} when the NumRel $\Omega(t)$ of the $\chi=-0.8$ model is used. For comparion the NumRel results are shown for both the $\chi=-0.8$ and $\chi=0$ models. It should be understood that the Newtonian and PN approximations depend sensitively on the choice of the coalescence time. With different choices, these curves could be made to fit the NumRel results significantly better well before the merger.

\bigskip

\subsubsection{Post-merger fit}


Shortly after the merger, the remnant, the single black hole that is formed, undergoes ringdown oscillations. As already pointed out in Sec.~\ref{sec:PhenFromNumRel}, for our $\chi=-0.8$ model this means a damped oscillation at complex QN frequency~\cite{BertiTables} $\omega=\omega_{\rm R}+\omega_{\rm I}\,i=(0.4427- 0.08675\,i)/M^{\rm rem}    =(0.4579-0.08975\,i)/M.$
The magnitude of the energy flux, $\dot{E}$, which varies as the square of these oscillations should have a time 
dependence proportional to $e^{-2\omega_{\rm I}t}$. This is tested in Fig.~\ref{fig:bothsides}. The plot shows $\dot{E}$ from NumRel compared with the QN exponential fall off; the magnitude is adjusted by eye for the best fit.

\begin{figure}
    \centering
    \includegraphics[width=.45\textwidth]{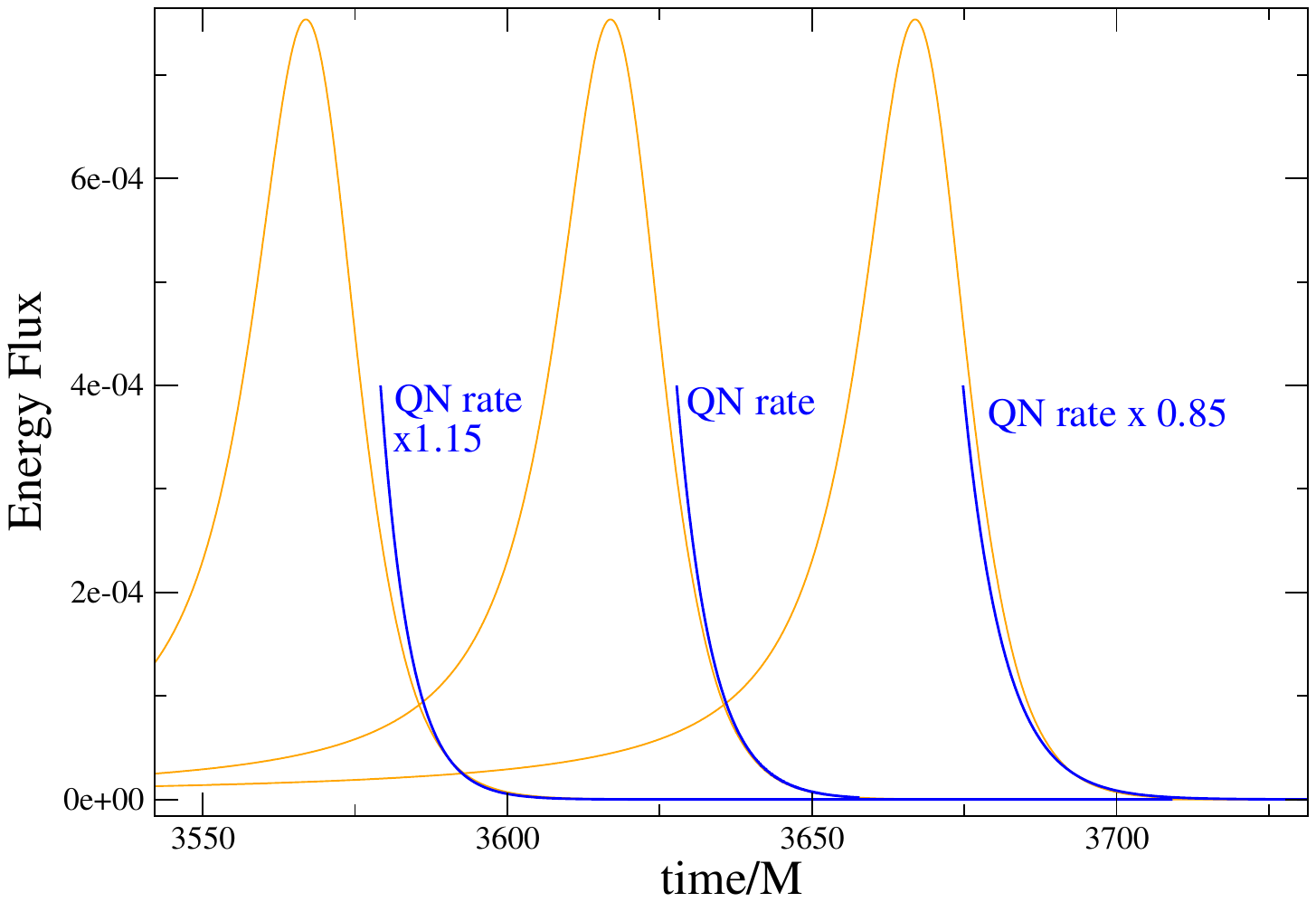}
    \caption{   $\dot{E}$ from NumRel compared with exponential fall-off in time. The middle plot shows the fall-off at the quasinormal damping rate for the remnant Kerr hole. The other plots show the sensitivity of the fit to changes in the fall-off rate.}
    \label{fig:QNcomparison}
\end{figure}

\begin{figure}
    \centering
    \includegraphics[width=.45\textwidth]{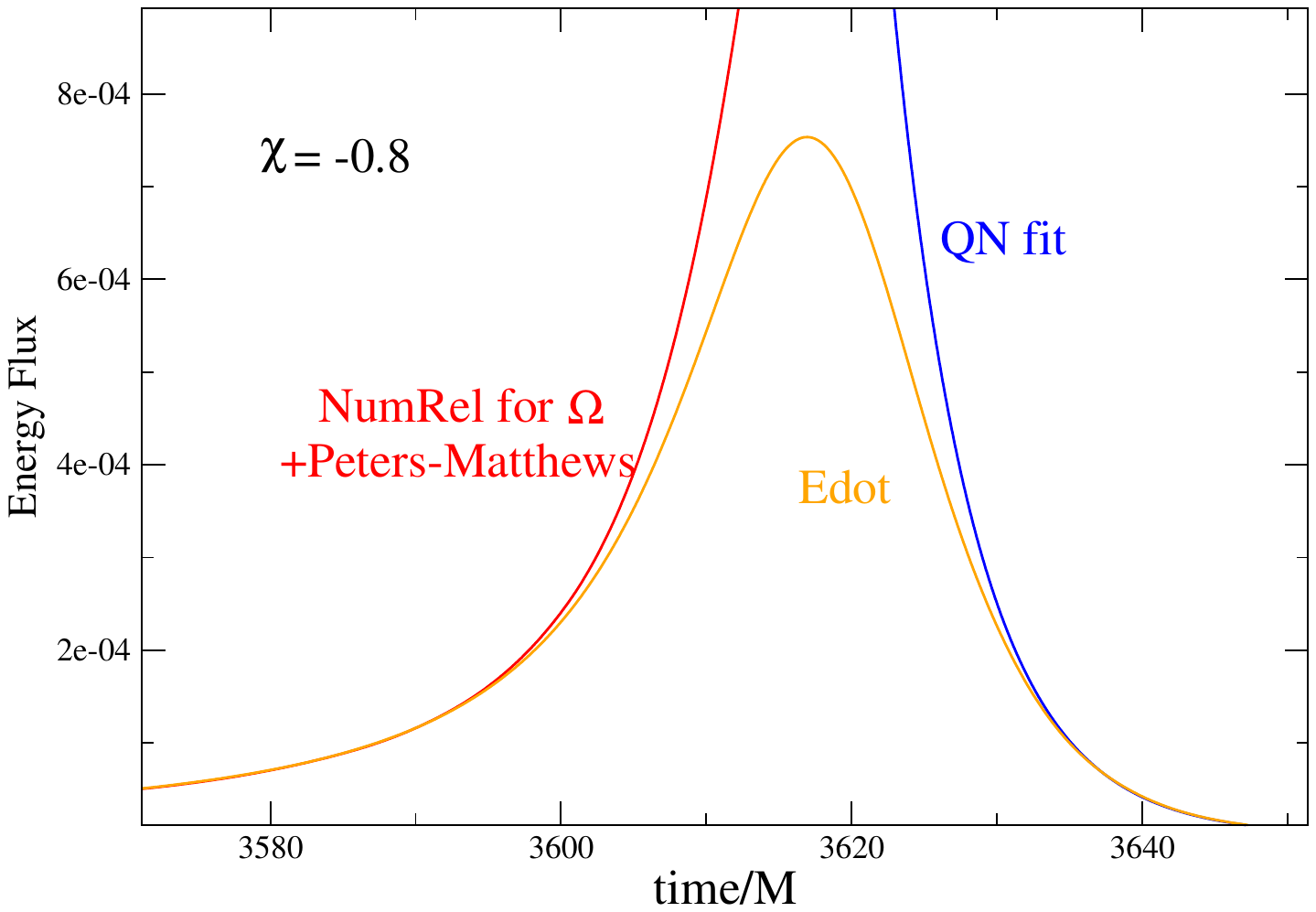}
    \caption{The approximation from Fig.~\ref{fig:preplunge} along with a fit proportional to $e^{-2\omega_{\rm I} t}$}.
    \label{fig:bothsides}
\end{figure}

One might worry that the fit of $\dot{E}$ to QN ringing is a demonstration only of exponential fall off, and is not specific to fall off at the QN rate. To address that worry we include, in Fig.~\ref{fig:QNcomparison}, time-shifted plots of $\dot{E}$ and display the best fit (by eye) to exponential fall at rates 15\% greater (on left) and 15\% less (on right) than the true QN fall-off rate. These modified fall-off rates clearly do not match as well as the QN rate 
for $t/M$ less than (shifted) 3635.

Figure~\ref{fig:bothsides} combines the pre-merger approximation of Fig.~\ref{fig:preplunge}
and the post-merger exponential fall-off of Fig.~\ref{fig:QNcomparison}.   Both the pre- and post-merger curves come from the basic physics of the BBH inspiral, not from NumRel. What is missing from a complete ``deconstruction" of the emission during the coalescence is a treatment of the middle, the merger. This would mean  the normalization of the post-merger curve, or -- equivalently -- the maximum value of the $\dot{E}$ curve that connects pre- and post-merger evolution. 
We turn now to estimating that value.

\section{Approximations for the peak $\dot{E}$ and the importance of the ``CLAP"}\label{sec:peakEdot}

In order to complete the picture started in Fig.~\ref{fig:bothsides} we must find a bridge between the pre-merger and post-merger curves; we must have a way of approximating the energy flux during the merger, the epoch that is both most important to the generation of GWs and most difficulty to analyze. This brings us to what is the most significant element of the current paper: the inclusion of an approximation appropriate to the epoch of the merger.

That approximation is a form of the close limit approximation~\cite{CLAP1,CLAPcomparison,CLAP2,CLAP3}, the CLAP, inspired by the ``Misner wormhole"~\cite{MisnerWormhole}. A more technical discussion is given in Appendix~\ref{app:CLAP}; here we want to emphasize the broader implications and usefulness of the CLAP class of approximations.

In short, a CLAP calculation starts with a way of giving solutions of the initial value equations of general relativity when two (or more) compact objects are close enough together that the  initial geometry can be considered to be a small deviation of the single remnant black hole that results from the merger of the objects. The deviations are treated as a linear perturbations of the black hole spacetime of the remnant black hole, so that Einstein's equations linearize  in these perturbations and are  relatively easy to evolve computationally. (This computational evolution is not considered NumRel.) That evolution shows QN ringing and hence allows us to link the strength of that ringing to the pre-merger features of the merging holes.

The original CLAP model entailed equal mass  unspinning BHs in a  momentarily stationary configuration~\cite{CLAP1,CLAPcomparison}. Subsequent work added momenta along the line of separation, for a head-on collision; momentum orthogonal to the line of separation, to represent angular momentum; spin of the premerger black holes; unequal masses~\cite{AndradePrice} and unequal masses with angular momentum~\cite{SopuertaYunesLaguna}.

Underlying these models, and especially appropriate to our use of CLAP, is the justification for the application. We are taking the CLAP initial conditions to stand in for the true conditions at the start of the merger. The true initial conditions involve the spacetime that results from the long inspiral of the component holes, up to the point at which the merger starts. Clearly there are differences in detail. The true conditions, for example, contain the relatively weak gravitational waves far from the merger location, in addition to the close-in spacetime details that will disappear inside the horizon that forms. Nevertheless, the CLAP conditions represent the {\it physical} conditions of the merger moment, and no incoming gravitational waves. 

The pudding in which the proof can be found for the substitution of CLAP conditions for true conditions 
is the agreement of CLAP with NumRel in the scenarios in which it has been tested. An example is found in Baker {\it et al.}~\cite{CLAP3}. The physical scenario is two equal mass BHs starting from rest at some initial separation $L$, with equal and opposite momentum $P$ parallel to the line of separation. For cases with $P\!\ll\!M$ the difference of the NumRel and the CLAP results in Fig.~2 of Baker {\it et al.}~\cite{CLAP3}  are too small to be resolved.

A secondary purpose of the current paper is to encourage further refinement of CLAP models, by showing how useful the CLAP can be as a supplement to NumRel. 

\begin{figure}
    \centering
    \includegraphics[width=.45\textwidth]{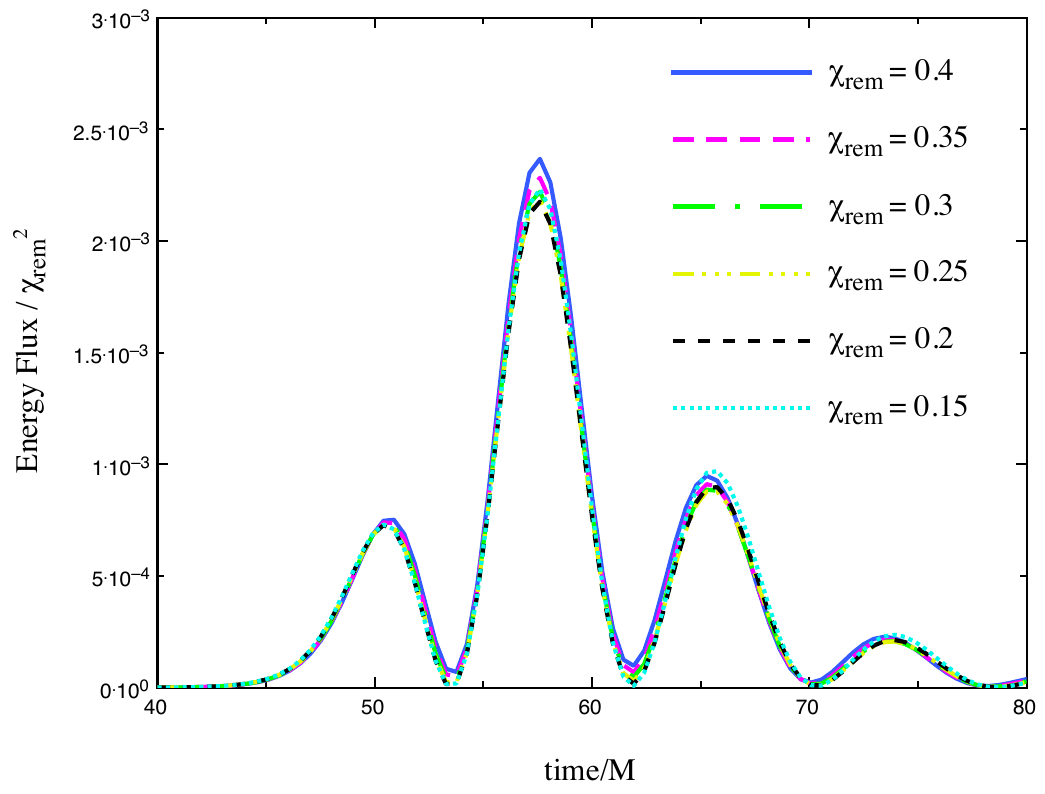}
    \caption{$\dot{E}/\chi_{\rm rem}^2$ from CLAP with data taken from Ref~\cite{CLAP2}. We approximate the ``CLAP index'' to be $2.3\times 10^{-3}$ from this figure.}
        \label{fig:eflux_clap}
\end{figure}

For the current paper we will only use the results generated in Ref.~\onlinecite{CLAP2}. Figure~\ref{fig:eflux_clap} shows a plot of $\dot{E}/\chi_{\rm rem}^2$ as a function of time generated by solving the Teukolsky equation for CLAP initial data that represents a binary with different values of $\chi_{\rm rem}$. The peak value of this flux data is 
seen to be well represented by
\begin{equation}\label{eq:CLAPindex}\Epeak \approx   
  2.3\times10^{-3}\chi_{\rm rem}^2\,.
\end{equation}

\begin{figure}
    \centering
    \includegraphics[width=.45\textwidth]{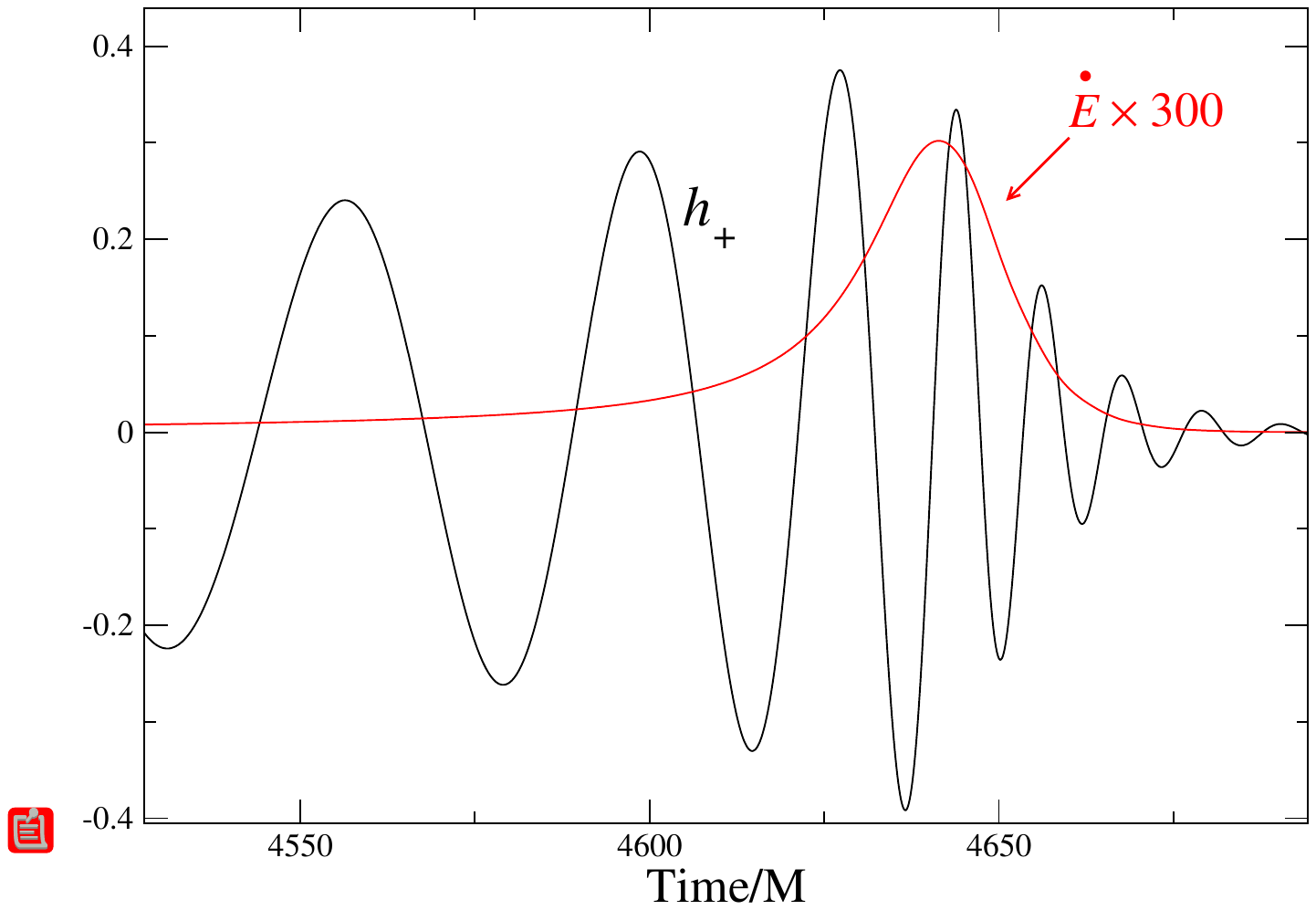}
    \caption{$\dot{E}\times300$ and the GW strain $h_+$.}
    \label{fig:NoSpinForPaper}
\end{figure}

For the $q=1$, unspinning model, Fig.~\ref{fig:NoSpinForPaper} shows the last few cycles of the quasicircular inspiral, the plunge and the onset of QN ringing. In order to apply Eq.~\eqref{eq:CLAPindex} we need to estimate 
the relevant value of $\chi$ entering into the QN ringing phase, hence we must estimate the mass and angular momentum. 

A good approximation is to use  $\Mrem$ and $\Jrem$, the values at the {\it end} of QN ringing. To make this choice we should check that a significant amount of mass and angular momentum is not radiated after the start of QN ringing. The nature of the problem appears in Fig.~\ref{fig:NoSpinForPaper}, which shows the merger era, a few cycles previous to it, and the QN ringing after it. Overlaid is a plot of the rate of energy loss to GW emission.

For our $\chi=-0.8$ model we saw, in Figs.~\ref{fig:QNcomparison} and \ref{fig:bothsides}, and in the discussion that followed, that ringdown started at $t/M=3635$. In a similar analysis for the $\chi=0$ model we find that the start of ringing is at $t/M=4652$. 
Figure~\ref{fig:NoSpinForPaper} suggests that the bulk of the energy emission is before $t/M=4652$. Integrating $\Edot$ and $\Jdot$ confirms this. 

\begin{figure}
    \centering
    \includegraphics[width=.45\textwidth]{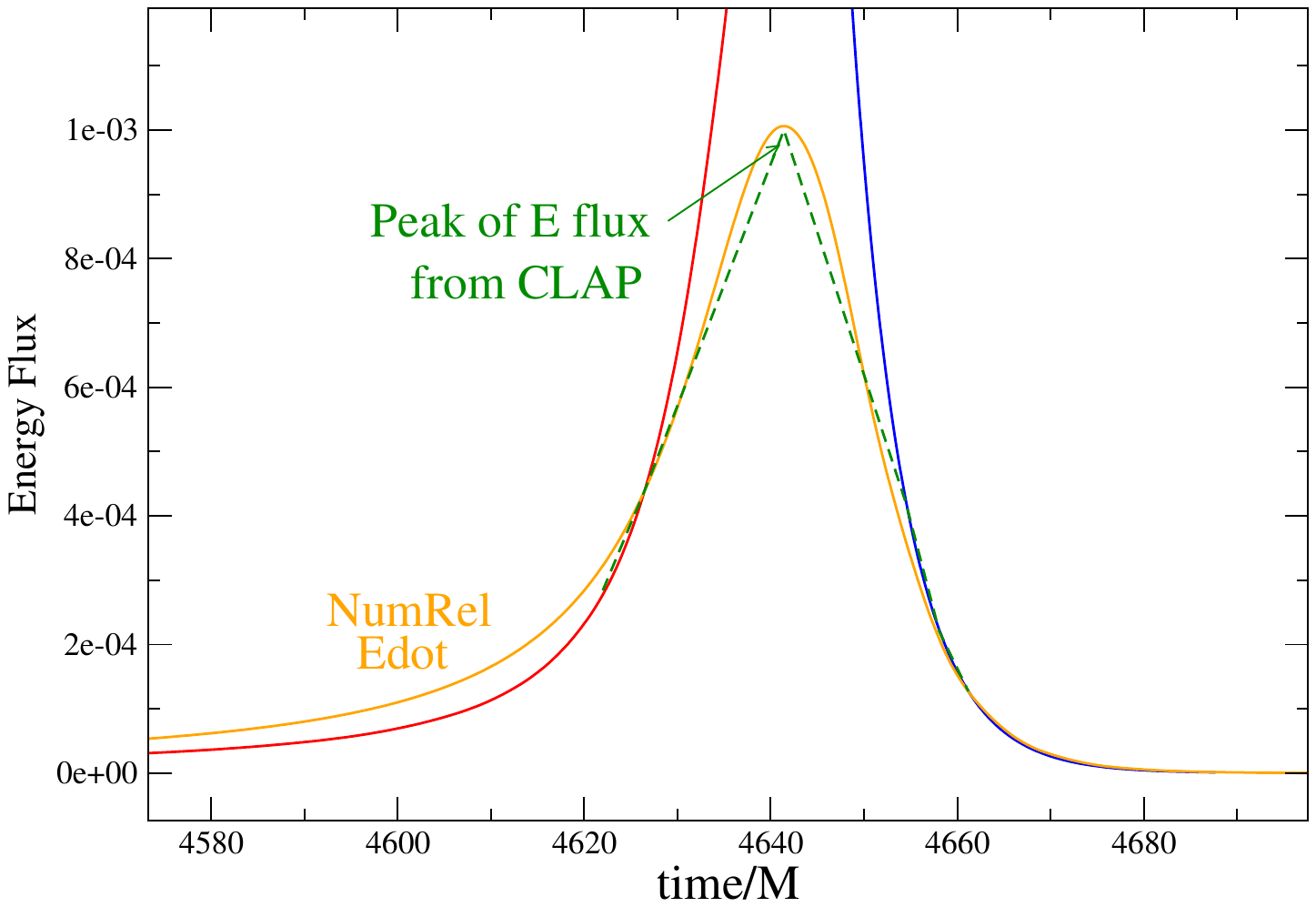}
    \caption{$\dot{E}$ for the $q=1$ spinless model. The CLAP peak is used to bridge the pre- and post-merger curves.}
    \label{fig:addingcap4}
\end{figure}

The remnant values for this $\chi=0$ model are $\Mrem=0.9516$, and $\Jrem=0.6216$. The mass radiated subsequent to  $t/M=4652$ is $0.012$ or about 1\% of $\Mrem$; the angular momentum radiated subsequent to 4652 is around 0.01, or 1.5\% of $\Jrem$. In view of the approximate nature of these calculations we 
will ignore these small amounts and use the remnant value $\chirem=\Jrem/\Mrem^2$ in Eq.~\eqref{eq:CLAPindex}, with the resulting prediction $\Epeak=1.084\times10^{-3}$. The NumRel value, exhibited in Fig.~\ref{fig:NoSpinForPaper} is $\Epeak=1.007\times10^{-3}$. The difference is less than 8\%. In Fig.~\ref{fig:addingcap4} this peak value is used to bridge the NumRel+quadrupole pre-merger and QN-fit post-merger curves for our $q=1$ spinless model.

Work is needed to extend this analysis to models with spin, and to models with unequal mass. If these models give good estimates of the NumRel value of $\Epeak$, it will go a long way to support the argument that CLAP initial data are sufficient for good estimates of the evolution of systems ending in a remnant black hole.

\section{An attempt to predict NumRel-independent coalescence}\label{sec:NumRelInd}

In what is above we have tried to understand the results of NumRel computations of BBH by using Newtonian physics (pre-merger orbital physics), linearized general relativity (quadrupole formula for GW emission), and perturbations of black hole spacetimes (QN oscillations and CLAP).  
We now ask about the reverse task: not to {\it understand} NumRel with the above techniques,  but to use these techniques to {\it predict} the outcome, or at least an approximation of the outcome of BBH coalescence. In particular, can we predict, independently of NumRel, the energy radiated and the angular momentum radiated? Can we predict the properties of the remnant, its mass and its spin? 

We start by reviewing what NumRel results were used in arriving at the three-part fit in Fig.~\ref{fig:addingcap4}. For both the ringdown complex frequency and for the application of a CLAP approximation, we used the remnant values $\Mrem$ and $\Jrem$ from NumRel. For the pre-merger fit we used $\Omega(t)$
from NumRel. We now discuss finding approximations for those remnant values without resorting to NumRel.

\subsubsection{Predicting the remnant angular momentum}\label{subsub:predictingJrem}
Our estimate of the remnant angular momentum $\Jrem$ is based on an extension to comparable mass holes of particle perturbation theory. We use the orbital angular momentum $L_K$ of a particle in equatorial orbit at the innermost stable orbit (`ISCO') of a Kerr hole. The way in which this result is extended from particles to comparable mass holes is spelled out in Appendix~\ref{app:BOAM}. The remnant angular momentum follows from adding the black hole spin angular momentum to  $L_K$.

In Table~\ref{tab:KerrInspiredJrem} we provide predictions of  $\Jrem$. The prediction uses the orbital angular momentum for this Kerr-extension, and a comparison with the NumRel result. 
Two predictions are listed. The first, prediction A, in column 4, uses the NumRel result for the remnant mass $\Mrem$ for each  BH. $\Mrem$ enters the prediction in two ways. First, it enters into the Kerr-extension, as shown in Appendix~\ref{app:BOAM}. Second, it enters in the calculation of spin angular momentum of the holes at the merger from the $\chi$ value. Prediction B is truly predictive. It assumes that the system mass remains unchanged at $M=1$, and that $\chi$ remains unchanged. (This is equivalent to $\Jspin$ remaining unchanged, since $M$ is unchanged.)

In Appendix~\ref{app:tide} we discuss the question of whether $\Jspin$, the spin angular momentum of one of the holes can change during inspiral and how this could come about through tidal interaction. The discussion in Appendix~\ref{app:tide} suggests that tidal interaction has a negligible effect, but that the question of the evolution of $\Jspin$ needs further study. Here we assume that $\Jspin$ does not change, and we take some comfort in the impressive agreement between our approximation for $\Jrem$, and the NumRel result.

The agreement of prediction A (column 4)  with the NumRel computation of NumRel is remarkably accurate for the range of models considered. The worst disagreement is less than 3\%. That column, however, is not strictly `predictive' since it uses as input the NumRel value of $\Mrem$. It is included primarily to show the reliability of the Kerr-extension for $L_K$, but it can be considered to have an additional purpose. If the NumRel-independent predictions for the coalescence are iterated, a value of $\Mrem$ can be found in the first iteration as explained in Subsection~\ref{susub:PredictingMrem} below, and that prediction can be used to improve the prediction for $\Jrem$ in the next iteration.

Prediction B (column 5) of $\Jrem$ in Table~\ref{tab:KerrInspiredJrem} gives the NumRel-independent `first iteration' prediction of $\Jrem$. Although the numerical agreement falls well short of that given by prediction A, it is tolerable for understanding trends and dependencies.

\begin{table}[h]
  \caption{Kerr inspired approximation for $\Jrem$. Column 4 (prediction A) is the predicted $\Jrem$ using the NumRel value of $\Mrem$ and the assumption of constant $\Jspin$. Column 5 (prediction B) uses  unchanging system mass $M=1$.  }\label{tab:KerrInspiredJrem}
$$
\begin{array}{|c|c|c|c|c|c|}\hline
  \chi_1,\chi_2 &\Mrem& L_K &\Jrem \mbox{A} &\Jrem \mbox{B}&\mbox{\!NumRel\!} \\ \hline
   -0.8 & 0.9666&0.7780& 0.4043& 0.4327  &0.3987 \\  
     -0.5&0.9622&0.7313 & 0.4999  &   0.5399 &0.4882\\  
   -0.3 &0.9586 &0.6973&0.5595 &   0.6088  &0.5443\\
   0&0.9516&0.6403&0.7071&0.6403&0.6216\\
  0.8 &0.9114& 0.4142&0.7465&   0.8987   &0.7538 \\  
     0.5 &0.9327&0.5201& 0.7376 &  0.8479    &0.7231 \\  
    0.3 &0.9418 &0.5734&0.7064&   0.7964  &0.6879\\
 \hline
\end{array}
$$
\end{table}

\subsubsection{Predicting $\Jrad$}\label{susub:PredictingMrem}

To make a NumRel-independent prediction of $\Jrad$, the angular momentum radiated during the BBH coalescence, we use the fact that only {\it orbital} angular momentum can couple directly to GWs. The models of coalescence start with relatively slow orbiting, so that the initial orbital angular momentum $\Lin$ for our $q=1$ models can be estimated from the initial angular velocity $\Omega_{\rm in}$ from the Newtonian relationship $\Lin=M^{5/3}\Omega_{\rm in}^{-1/3}\!\!/4$. The difference between $\Lin$ and $L_K$ derived in Appendix~\ref{app:BOAM}, and in subsection~\ref{subsub:predictingJrem}, can be ascribed to angular momentum carried off in GWs. The results of the subtraction, and the comparison with the NumRel value of $\Jrad$ is shown in Table~\ref{tab:LinMinusLk} for our $q=1$ models.

\begin{table}[h]
  \caption{NumRel-independent prediction of $\Jrad$ compared with NumRel result. The first column is the Newtonian estimate of the initial orbital angular momentum; the second column is the Kerr-inspired value derived in Appendix~\ref{app:BOAM}. }\label{tab:LinMinusLk}
\vspace{-.2in}
$$
\begin{array}{|c|c|c|c|c|}\hline
  \chi_1,\chi_2& \Lin &L_K&\Lin\!\!-\!\!L_K &\mbox{NumRel} \\ \hline
   -0.8&1.014 &0.7780&   0.236  &0.348\\  
    -0.5& 1.020 &   0.7313  & 0.289     &0.391\\  
  -0.3& 1.030 &    0.6973  &   0.333   &0.425\\
0&1.086&0.6403&0.482&0.482\\
  +0.8& 1.025&  0.4142 &     0.611 &0.695\\  
    +0.5& 1.020  &  0.5201  &   0.500    &0.606 \\  
    +0.3& 0.994   &  0.5734  &     0.421   &0.550 \\
 \hline
\end{array}
$$
\vspace{-.0in}
\end{table}

The agreement of the NumRel value (column 5) with the NumRel-independent prediction (column 4) is not highly accurate. This is almost certainly caused by the error in using a Newtonian estimate for $\Lin$. The initial orbital velocity is around 0.12\,c, which implies moderate errors in the Newtonian estimate, errors that are, of course, magnified in the subtraction given in column 4.

\subsubsection{Predicting $\Mrem$}\label{susub:PredictingMrem}
Estimating the NumRel-independent value of $\Mrem$ is equivalent to an estimate of the GW energy radiated during the BBH. This requires an estimate of the energy flux $\Edot(t)$. We have discussed in Sec.~\ref{sec:peakEdot} how the CLAP approach gives a good estimate of the peak of this flux if $\Jrem$ is known, and we have seen in subsection~\ref{subsub:predictingJrem} how to make a NumRel estimate of $\Jrem$. 
We shall not discuss $\Jrem$ further here since there is a much more important difficulty in finding a NumRel-independent estimate of $\Mrem$. In addition to $\Epeak$  
 we  need to know the period of time during which there is strong GW energy flux. The nature of the problem can be seen in Fig.~\ref{fig:NoSpinForPaper}, and we use that figure as the basis for our estimates.

In that figure the strong emission occurs at around $t/M=4640$
 The period of strong emission extends for only around a single GW cycle, and it occurs at a time at which the $\Omega(t)$ curve is bending rapidly upward. From the Newtonian or post-Newtonian curves in Fig.~\ref{fig:PN2rvsdNumRel} we infer a value of $\Omega$ between 0.1 and 0.15, implying a time for a GW cycle between 42 and 62. 
 From Table~\ref{tab:KerrInspiredJrem}
we take $\Jrem=0.7$ for $\chi=0$ to be the value of $\chi$. Along with the estimate of $\Epeak$ from Eq.~\eqref{eq:CLAPindex}, this means total radiated energy be 0.05 and 0.07, in very rough agreement with the  NumRel value 0.04.


\section{Conclusions}

We have confirmed the reasonable accuracy of approximations in the pre-merger and post-merger stages of binary black hole coalescence. Our most important addition to the phenomenology of the coalescence has been to give preliminary evidence for dealing with the merger itself by using the  close-limit approximation (CLAP). 

The CLAP had been used previously, with excellent results, for BH interactions starting with the participating BHs close to each other, and constrained to a region that will evolve 
to be inside a developing horizon. What we have found in this paper is that the CLAP 
has a much broader applicability. It can be used in scenarios in which the ``initial" data are the three-geometry and extrinsic curvature that are inferred from earlier evolution with approximation methods appropriate to that earlier evolution. For reasonably accurate predictions of merger GW radiation, all that is needed from the earlier evolution is the phenomenological parameters for the input to the CLAP.

With this new addition to the package of methods and insights we have given a reasonably good phenomenological picture of all stages of binary black hole coalescence, at least for a limited set of BBH configurations. Perhaps more important, we have shown that with the set of tools exhibited, it is possible to make NumRel-independent predictions of GW radiation that may be adequate for many astrophysical applications. Confidence in this approach awaits a study of a much broader class of models than is considered in this paper, a paper in which the focus has been the core methods.

Along with the above we have identified a few specific areas that will require further study and have found a few details worthy of notice. Among them:

\smallskip
$\bullet$ In this paper we have compared NumRel and CLAP only for BBH with unspinning BHs. CLAP solutions exist\cite{GauravPaper} for spinning BHs. Comparisons must be made, in the same manner as in the current paper, of those CLAP solutions and NumRel.

\smallskip
$\bullet$ We have found that the quadrupole approximation, for motion at a given angular velocity [See Eq.~\eqref{eq:PetersMatthews2}] remains accurate even in the late coalescence up to the approach of very brief period of merger. The error in a purely Newtonian approximation to that approach resides in the Newtonian inaccuracy in estimating the evolution of the angular velocity, especially for BHs with spin.

\bigskip
{\em Acknowledgments:}  
R.B. and G.K. acknowledge support from NSF Grants No. PHY-2307236 and DMS-2309609. All computations were performed on the UMass-URI UNITY HPC/AI cluster at the Massachusetts Green High-Performance Computing Center (MGHPCC).

\appendix

\section{Close-Limit Approximation (CLAP)}\label{app:CLAP}

It is useful, especially for computational work, to separate the full set of Einstein's equations into an ``inital value" set and a set of evolution equations that solve for the time development of the solution to the Einstein equations. In the ADM formalism~\cite{chap21MTW} that is typically used in NumRel there is a ``Hamiltonian constraint" on the initial 3-geometry in the form of a single scalar-like relationship between the Ricci scalar $^{(3)}R$ of the initial spacetime slice~\cite{conflat} and the extrinsic curvature $K_{ab}$ of that slice. 

The initial value equations also include a ``momentum constraint," a 3-dimensional divergence condition that must be satisfied by $K_{ab}$. The first application of the CLAP approximation was for the case $K_{ab}=0$, so that the spacetime that evolves is time symmetric about the initial slice. That first application represented initial data for two black holes~\cite{waszismean} separated by an adjustable distance. If that separation is chosen small enough relative to the size of the throats, then at a distance from a midpoint that is several times the separation, the 3-geometry is that of a single throat with small perturbations. The details at lesser distances from a midpoint will affect the evolution only inside the horizon that evolves. It is therefore only the perturbations further out that will be of interest, and these can be treated with the techniques of black hole perturbation theory, in particular the Regge-Wheeler-Zerilli~\cite{RWZ} equation if the single black hole is nonspinning, and the Teukolsky equation~\cite{Teukolsky} if it is spinning.

After the demonstration of the usefulness and accuracy of the $K_{ab}=0$ first CLAP model, the two-throat structure of the initial 3-geometry was maintained, but the extrinsic curvature was used to represent starting spacelike slices that were not moments of time symmetry. In particular, initial momentum parallel to the line of separation of the two throats~\cite{CLAP3} was used to represent black holes moving toward each other at the initial moment; momentum perpendicular to that line represented black holes moving past each other, or in binary orbits~\cite{CLAP2}. The intrinsic spin of the individual holes could also be embedded in the initial form of $K_{ab}$ and used in a CLAP application.~\cite{GauravPaper}

\section{Kerr-Inspired Binary Orbital Momentum} \label{app:BOAM} 

Analytic expression of binary orbital angular momentum (BOAM) is not readily available in case of two black holes orbiting each other. Ref.~\onlinecite{BPK} provides an analytical expression for BOAM starting from point-particle approach. Orbital angular momentum of a particle of mass $\mu$ in a circular orbit or radius $r$ around a Kerr black hole of mass $M$ is given by,

 \begin{equation}\label{eq:Lk}
  \frac{\Lk}{\mu}=\frac{\pm M^{1/2}\left(r^2 \mp 2aM^{1/2}r^{1/2} + a^2\right)}{r^{3/4}\left(r^{3/2} - 3Mr^{1/2}\pm 2aM^{1/2}\right)^{1/2}}
\end{equation}
where $\mu << M$, and where ‘$+$’ and ‘$-$’ signs represent the prograde and retrograde motion respectively. One can calculate the angular momentum of hole 2 denoted here as $L^{\rm Kerr}_{2}$, by setting $\mu = m_2$, $M = m_1$, and $a = m_1 \chi_1$. Likewise, $L^{\rm Kerr}_{1}$, the orbital angular momentum of hole 1 in the background of hole 2, can be calculated by setting appropriate mass and spin parameters.

The above expression treats one black hole as stationary, and the other as following geodesic motion. For comparable mass holes, however,  holes orbits their common barycenter. To account for this $L^{\rm Kerr}_{1}$ and $L^{\rm Kerr}_{2}$ is modified to obtain binary orbital angular momentum $L^{\rm bin}$,

\begin{equation}
    L^{\rm bin} = \Gamma_1 L^{\rm Kerr}_1 + \Gamma_2 L^{\rm Kerr}_2\,.
\end{equation}
Here $\Gamma_1 = (1+q)^{-3/2}$, converts the angular momentum of hole 1 to barycenter value in Newtonian physics, as explained in Ref.~\onlinecite{BPK}. Similarly, $\Gamma_2 = (1+1/q)^{-3/2}$ makes the conversion for hole 2. 

\section{Tidal transfer of angular momentum}\label{app:tide} 
Particle perturbation (PP) studies~\cite{Poisson70,tidalrefs} have shown that nearby objects can raise a ``hill" on the horizon of a hole. If the hole is rotating, the hill lags the source of the perturbation much as a tidal hill of a rotating viscous star lags. The interaction of that perturbing hill and the tidal field produce a torque that slows the rotation of the hole. 

Extensions of PP to binaries of comparable mass BHs has been seen to be remarkably effective, e.g., in Appendix~\ref{app:BOAM}. Here we use such an extension simply to get a rough estimate. We infer that the tidal torque is proportional to the tidal strength (mass of perturbing object $M_{\rm pert}$ divided by the cube of ${Sep}$ the distance to that object). The extent to which a tidal ``hill" is raised is also proportional to this tidal strength. The tidal torque is therefore proportional to $\left(M_{\rm pert}/{Sep}^3\right)^2$. To apply this to a rough estimate of torque in a comparable-mass binary we take $M_{\rm pert}=M$ and ${Sep}=R$, and use the Keplerian relation of $M,R$ and $\Omega$ to find that the torque is proportional to $\Omega^4$.

The particle perturbation studies, and the stellar analogy suggest that the tidal torque is proportional to the relative speed of the perturbing source and the ``viscous" fluid of the affected star. We will follow Poisson~\cite{Poisson80IVB} in using the horizon angular velocity $\OmegaH$ to represent the rotation rate of the perturbed horizon. 

A reasonable guess at a tidal torque expression extendable to comparable mass holes is 
\begin{equation}\label{eq:tidaltorque}
    \mbox{tidal torque}=\frac{dJ}{dt}=M^2\frac{d\chi}{dt} =\beta M^6|\Omega-\OmegaH|
\Omega^4\,,
\end{equation}
where $\beta$ is a constant of order unity. The factor of $M^6$ on the right, which is necessary for dimensional consistency, arises from the integration of the tidal interaction over the affected BH.

In this approach, the tidal transfer of angular momentum may become significant in either of two ways. 
Due to the extreme sensitivity (inverse sixth power) to separation, tidal transfer may be significant in the short time, just before merger, during which the separation is small, but the two BHs are as yet not merged. It may also be significant, though weak, for the long time of the slow decay of the orbit.

\begin{figure}
    \centering
    \includegraphics[width=.4\textwidth]{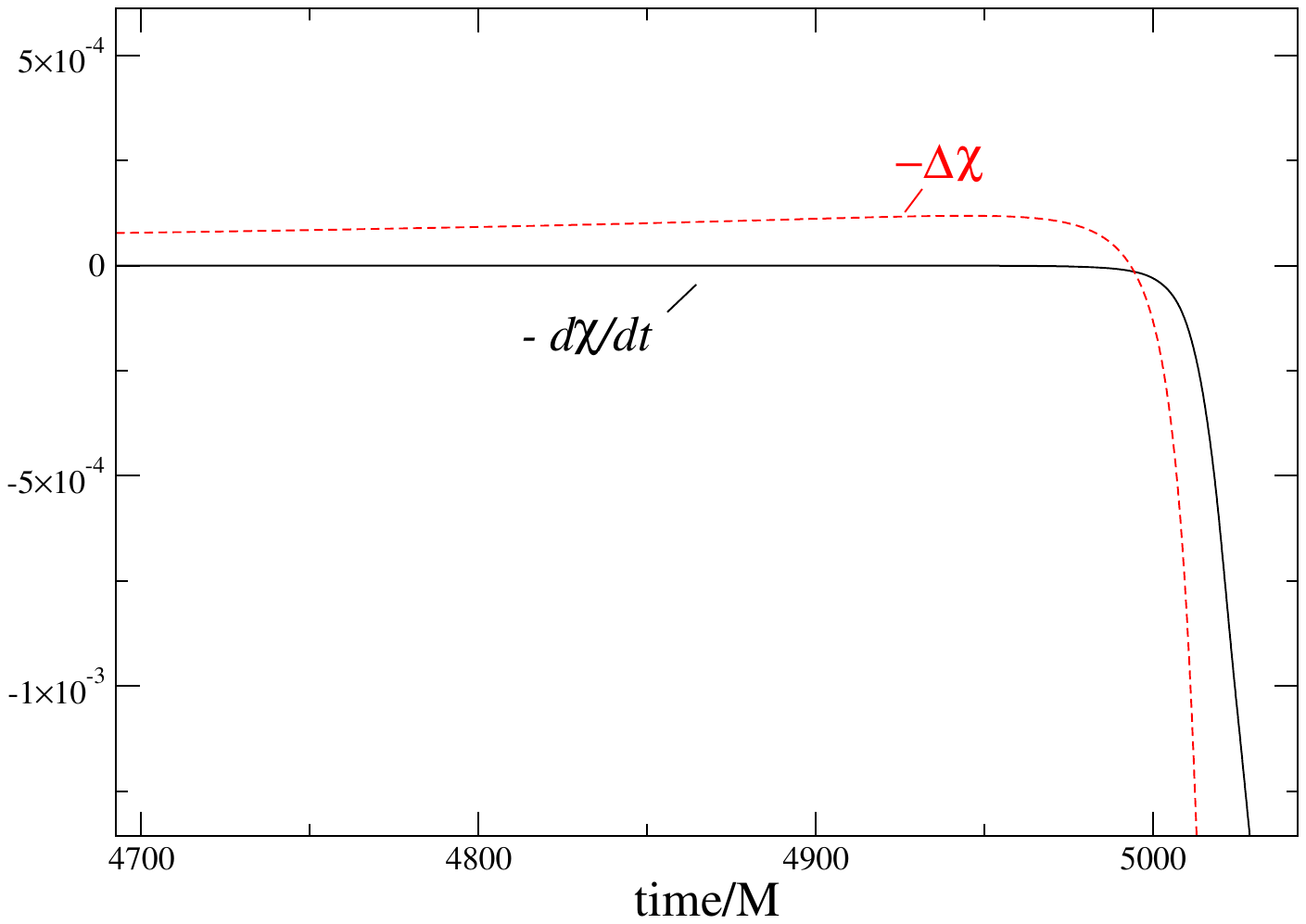}
    \caption{Tidal transfer of angular momentum for the symmetric BH coalescence model with $q=1$ and $\chi=0.3$. The black curve plots $\Omega^4(\OmegaH-\Omega)$, representing the tidal torque.
     The red (dashed) curve shows the time integral of that expression from $t=0$ in the model SXS:BBH0164~\cite{SXScatalog} at which $M\Omega=1.590\times10^{-2}$.} 
    \label{fig:plus03}
\end{figure}
Figure~\ref{fig:plus03}, for the symmetric model with $\chi=0.3$, illustrates the problem of inferring the tidal transfer. Even over thousands of time units leading up to the merger, the total effect on $\chi$ is negligible. As the merger is reached (at around $t/M=5000$, where the curve $d\chi/dt$ curve dips precipitously) the rate increases dramatically, but the net change in $\chi$ is negligible, unless we consider times well into the merger, at which the concept of tidal transfer does not really apply.   We have looked at similar analyses of $\Delta\chi$ for  all the $q=1$ models considered in this paper. All such analyses are similar to what is shown in Fig.~\ref{fig:plus03}; in all cases tidal transfer of angular momentum is negligible until it is meaningless.

\pagebreak


\begin{thebibliography}{11}

\bibitem{BaumShap}
Thomas W. Baumgarte and Stuart L. Shapiro, {\em Numerical Relativity: Solving Einstein's Equations on the Computer}, 
Cambridge University Press \url{https://doi.org/10.1017/CBO9781139193344} (2010).

\bibitem{SXScatalog}
SXS Waveform Catalog \url{https://data.black-holes.org/waveforms/catalog.html}.

\bibitem{SXSpaper}
Michael Boyle {\it et al.}, Class.~Quant.~Grav. {\bf 36}, 195006 (2019); Abdul Mroue {\it et al.} Phys. Rev. Lett. {\bf 111}, 241104 (2013).

\bibitem{vocab}	
Eanna E. Flanagan and Scott A. Hughes, Phys. Rev. D {\bf 57} 4535 (1998).

\bibitem{EdotVsChi}
Reinaldo Gleiser, Gaurav Khanna, Richard Price, Jorge Pullin, New J. Phys. {\bf 2}, 3 (2000).

\bibitem{CLAP1}
Richard H. Price and Jorge Pullin, Phys. Rev. Lett. {\bf 72}, 3297 (1994).

\bibitem{CLAPcomparison}
Peter Anninos, Richard H. Price, Jorge Pullin, Ed Seidel, Wai-Mo Suen, Phys. Rev. D {\bf 52}, 4462 (1995).

\bibitem{CLAP2}
Gaurav Khanna, John Baker, Reinaldo Gleiser, Pablo Laguna, Carlos Nicasio, Hans-Peter Nollert, Richard Price, Jorge Pullin, Phys. Rev. Lett. {\bf 83}, 3581 (1999) 

\bibitem{CLAP3} 
John Baker, Andrew Abrahams, Peter Anninos, Steve Brandt, Richard Price, Jorge Pullin and Edward Seidel, Phys.~Rev.~D {\bf 55}, 829 (1997).

\bibitem{AndradePrice}
Zeferino Andrade and Richard~H.~Price, Phys. Rev. D {\bf} 56, 6336 (1997).

\bibitem{SopuertaYunesLaguna}
Carlos F. Sopuerta, Nicolas Yunes, Pablo Laguna, Phys. Rev. D {\bf 74} 124010 (2006).


\bibitem{Poisson70}
Eric Poisson, Phys. Rev. D {\bf 70}, 084044 (2004).

\bibitem{Poisson80}
Eric Poisson, Phys. Rev. D {\bf 80}, 064029 (2009).
\bibitem{PetersMatthews} 
P. C. Peters and J.~Matthews, Phys.~Rev.~131,435 (1963).

\bibitem{MTW}
Charles~W. Misner, Kip~S. Thorne, and John~A. Wheeler, {\em Gravitation} (W. H.  Freeman, 
San Francisco, 1973).

\bibitem{MTWquadform} 
ibid, Sec.36.6.

\bibitem{BlanchetEtAl}
Luc Blanchet, Bala R. Iyer, Clifford M. Will, Alan G. Wiseman, Class. Quant. Grav. {\bf 13}, 575 (1996) (equation 8).

\bibitem{BertiTables} 
Emanuele Berti, Vitor Cardoso, Andrei O. Starinets, Class. Quantum Grav. {\bf 26}, 163001 (2009).

\bibitem{MisnerWormhole} 
Charles W. Misner, Phys.~Rev. {\bf118}, 1110 (1960)


\bibitem{GauravPaper} 
Gaurav Khanna, Phys. Rev. D {\bf 63}, 124007, (2001).

\bibitem{chap21MTW} 
Chapter 21 of Ref.~\onlinecite{MTW}  


\bibitem{conflat} In many NumRel models and in all CLAP approximations the initial geometry is taken to be conformally flat, so that solving the Hamiltonian constraint involves only solving for the conformal factor.

\bibitem{waszismean} If a throat of this type is isolated  it evolves  into a single black hole.


\bibitem{RWZ} 
Tullio Regge and John A. Wheeler, Phys. Rev. {\bf 108}, 1063 (1957); F. J. Zerilli, Phys. Rev. Lett. {\bf 24}, 737 (1970).


\bibitem{Teukolsky} 
Saul Teukolsky, Ap. J. {\bf 185}, 635 (1973).


\bibitem{BPK} 
Ritesh Bachhar, Richard H. Price, and Gaurav Khanna, Phys. Rev. D {\bf 108}, 064019 (2023).

\bibitem{tidalrefs} 
Eric Poisson, Phys. Rev. D {\bf 80}, 064029 (2009); Richard H. Price and Kip S. Thorne, Phys. Rev. D {\bf 33}, 915 (1986); Yotam Sherf,  Phys. Rev. D {\bf 103}, 104003 (2021); James B. Hartle, Phys. Rev. D {\bf 8}, 1010 (1973); James B. Hartle, Phys. Rev. D {\bf 9}, 2749 (1974); Stephen W. Hawking and James B. Hartle, Commun. Math. Phys. {\bf 27} 283 (1972).
\bibitem{Poisson80IVB} 
Sec.IVB of Ref.~\onlinecite{Poisson80}










\bibitem{fourthroot} More correctly: It is the fourth root of the conformal factor that satisfies the Laplace or Poisson equation.






\end{thebibliography}
\end{document}